\documentclass[5p]{elsarticle}
\usepackage{booktabs,shortvrb}
\usepackage{amsmath,amsfonts,amssymb}
\usepackage{booktabs,url}
\newtheorem{Theorem}{Theorem}
\newtheorem{Definition}[Theorem]{Definition}
\makeatletter
    \def\ps@pprintTitle{\let\@oddhead\@empty\let\@evenhead\@empty\let\@oddfoot\@empty\let\@evenfoot\@oddfoot }
\makeatother

\begin{document}
\begin{frontmatter}
\title{Sensitive Ants for  Denial Jamming Attack\\ on Wireless Sensor Network}
\author[1]{Camelia-M. Pintea}\ead{cmpintea@yahoo.com}
\author[1]{Petrica C. Pop}\ead{petrica.pop@cunbm.utcluj.ro}
\address[1]{Technical University Cluj Napoca, North University Center Baia Mare, Romania}
\begin{abstract}
A new defense mechanism for different jamming attack on {\it Wireless Sensor Network (WSN)} based on ant system it is introduced. The artificial sensitive ants react on network attacks in particular based on their sensitivity level. The information is re-directed from the attacked node to its appropriate destination node. It is analyzed how are detected and isolated the jamming attacks with mobile agents in general and in particular with the newly ant-based sensitive approach. 
\end{abstract}
\end{frontmatter}
\section{Introduction}
 Jamming attacks on wireless networks, special cases of {\it Denial of Service (DoS)} are  the attacks that disturb the transceivers’ operations on wireless  networks~\cite{adamy04}. A {\it Radio Frequency (RF)} signal emitted by a jammer corresponds to the ‘useless’ information received by the sensor nodes of a network.  

Nowadays are several techniques used to reduce the effect of jamming attacks in wireless networks. In \cite{yujin12} is proposed a traffic rerouting scheme for {\it Wireless Mesh Network (WMN)}. There are determined multiple candidates of a detour path which are physically disjoint. In a stochastic way, is selected just one candidate path as a detour path to distribute traffic flows on different detour paths. The mechanism on packet delivery ratio and end-to-end delay is improved when compared with a conventional scheme.

Securing {\it WSN}s against jamming attacks is a very  important issue. Several security schemes proposed in the {\it WSN} literature are categorized in: {\it detection techniques, proactive countermeasures, reactive countermeasures} and  {\it mobile agent-based countermeasures}. The advantages and disadvantages of each method  are described in \cite{Mpitziopoulos}. Our current interest is in {\it mobile agent-based solutions}.

Artificial intelligence has today a great impact in all computational models. Multi-agent systems are used for solving difficult Artificial Intelligence problems~\cite{wooldridge,Iantovics}. Multi-agents characteristics includes organization, communication,  negotiation, coordination,  learning, dependability, learning and cooperation~\cite{his,wooldridge,stoean}. Other features are their knowledge \cite{Popescu} and their actual/future relation between self awareness and intelligence. There are also specific multi-agents with their particular properties, as robots-agents~\cite{pintea2010,pinteacisis13}. One of the  commonly paradigm used with {\it MAS} systems is the pheromone, in particular cases artificial ants pheromone. Ant-based models are some of the most successfully nowadays techniques used to solve complex problems~\cite{crisan07,crisan,dorigo97,pintea2010,Reihaneh,stoean2010}.
  
Our goal is to improve the already existing ant systems on solving jamming attacks on {\it WSN} using the ants sensitivity feature. The second section describes the already known mechanisms of Jamming Attack on Wireless Sensor Network and the particular unjamming techniques based on Artificial Intelligence. The next section is about ants sensitivity. The section includes also the newly introduced sensitive ant-model for detecting and isolated jamming in a {\it WSN}. Several discussions about the new methods and the future works concludes the paper.

\section{About Jamming Attack on Wireless Sensor Network using Artificial Intelligence}
The current section describes the main concepts and the software already implemented on Jamming Attack on Wireless Sensor Network including the {\it Artificial Intelligence} models.
\subsection{Jamming Attack on Wireless Sensor Network}
Jamming attacks are particular cases of Denial of Service (DoS) attacks\cite{wood02}. The main concepts related to this domain follows. At first several considerations on Wireless Sensor Network (WSN).
\begin{Definition}
A Wireless Sensor Network (WSN) consists of hundreds/thousands of sensor nodes randomly deployed in the field forming an infrastructure-less network.
\end{Definition}
\begin{Definition}A sensor node of a {\it WSN} collects data and routes it back to the Processing Element (PE) via ad-hoc connections with neighbor sensor nodes.
\end{Definition}
\begin{Definition}
A Denial of Service attack is any event that diminishes or eliminates a network’s capacity to perform its expected function.
\end{Definition}  		
\begin{Definition}
Jamming is defined as the emission of radio signals aiming
at disturbing the transceivers’ operation.
\end{Definition}
There are differences between jamming and radio frequency interference.
\begin{itemize}
\item the jamming attack is {\it intentional} and {\it against a specific
target};
\item the radio frequency interference is unintentional, as a result of nearby transmitters, transmitting in the same or very close frequencies levels.
\end{itemize} 
An example of radio frequency interference  is the coexistence of multiple wireless networks on the same area with the same frequency channel~\cite{Mpitziopoulos}.
\begin{Definition}
Noise is considered the undesirable accidental fluctuation of electromagnetic spectrum, collected by an antenna. 
\end{Definition}
\begin{Definition}
The Signal-to-Noise Ratio is 
$$SNR= \frac{P_{signal}}{P_{noise}} $$
\noindent where $P$ is the average power.
\end{Definition}
A jamming attack can be considered {\it effective} if the {\it Signal-to-Noise Ratio(SNR)} is less than one ($SNR<1$).

There are several jamming techniques~\cite{Mpitziopoulos}: Spot Jamming, Sweep Jamming, Barrage Jamming and Deceptive Jamming.  

\begin{Definition}
‘Jammer’ refers to the equipment and its capabilities that are exploited by the adversaries to achieve their goal. 
\end{Definition}
\noindent There are several types of jammers, from simple transmitter or jamming stations with special equipment, used against wireless networks.\cite{xu05} 
\begin{itemize}
\item the constant jammer - emitting totally random continuous radio signals; {\it target}: keeping the {\it WSN}s channel busy; disrupting nodes’ communication; causing interference to nodes that have already commenced data transfers and corrupt their packets.
\item the deceptive jammer;
\item the random jammer - sleeps for a random time and jams for a random time;
\item the reactive jammer - in case of activity in a {\it WSN} immediately sends out a random signal to collide with the existing signal on the channel. 
\end{itemize}
\section{Sensitive Ant-based Technique for Jamming Attack on Wireless Sensor Network}
\noindent The current section describes the concept of sensitive ants proposed in \cite{chira07,pintea2010}. Effective metaheuristics for complex problems, as large scale routing problems (e.g. the {\it Generalized Traveling Salesman Problem}) based on sensitive ants are illustrated in \cite{chira07,his,pintea2010}. Several concepts are defined and described further.  
\begin{Definition}
Sensitive ants refers to artificial ants with a Pheromone Sensitivity Level (PSL) expressed by a real number in the unit interval $0\leq PSL \leq 1$.
\end{Definition}
\begin{Definition}
A pheromone blind ant is an ant completely ignoring stigmergic information, with the smallest PSL value, zero. 
\end{Definition}
\begin{Definition}
A maximum pheromone sensitivity ant has the highest possible PSL value, one. 
\end{Definition}
\begin{Definition}
The sensitive-explorer ants have small \\Pheromone Sensitivity Level values indicating that they normally choose very high pheromone level moves.
\end{Definition}
\begin{Definition}
The sensitive-exploiter ants have high \\Pheromone Sensitivity Level values indicating that they normally choose any pheromone marked move.
\end{Definition}

\noindent The sensitive-explorer ants are also called small PSL-ants, hPSL and sensitive-exploiter ants are called high PSL-ants, hPSL. They intensively exploit the promising search regions already identified by the sensitive-explorer ants. For some particular problems the sensitivity level of hPSL ants have been considered to be distributed in the interval (0.5, 1) and for the sPSL ants the sensitivity level in the interval (0, 0.5).

Based on the already described notions it is introduced a new ant-based concept with sensitivity feature. The {\it Sensitive Ant Algorithm for Denial Jamming Attack on Wireless Sensor Network} is further  called {\it Sensitive Ant Denial Jamming} on {\it WSN} algorithm.

As we know, not all ants react in the same way to pheromone trails and their sensitivity levels are different there are used several groups of ants with different levels of sensitivity. For an easy implementation are used just two groups-colonies.

 In \cite{Muraleedharan} is firstly introduced an ant system for jamming attack detection on {\it WSN}. The performance of the ant system is given by the node spacing and several parameters: $Q$ an arbitrary parameter, $\rho$ trail memory, $\alpha$ power applied to the pheromones in probability function and $\beta$ power of the distance in probability function.  

It is considered a {\it WSN} in a two dimensional Euclidean space. There are several key elements of AS for keeping the network robust and de-centralized. One is the information on the resource availability on every node used to predict the link for the ant’s next visit. 

\noindent Other key elements are the pheromones intensity and dissipate energy of ants as they traverse the nodes based on path probabilities. The key factor for making decisions in \cite{Muraleedharan} is the transition probability (\ref{eq:1}). In the newly {\it Sensitive Ant Denial Jamming} on {\it WSN} only the ants with small pheromone level are using this probability.

\begin{equation}\label{eq:1} 
P_{ij}=\frac{(\varphi_{ij}\cdot\eta_{ij})^\alpha\cdot(\frac{1}{D_{ij}})^\beta}{\sum_k (\varphi_{ik}\cdot\eta_{ik})^\alpha\cdot(\frac{1}{D_{ik}})^\beta}
\end{equation}
\noindent where $\eta_{ij}$ is the normalized value (\ref{eq:2}) of Hop, $H_{ij}$, Energy, $E_{ij}$, Bit Error Rate, $B_{ij}$, Signal to Noise ratio, $
SNR_{ij}$, Packet Delivery, $Pd_{ij}$ and Packet Loss, $Pl_{ij}$~\cite{Liang}.  
\begin{equation}\label{eq:2} 
\eta_{ij}=H_{ij}\cdot E_{ij}\cdot B_{ij}\cdot SNR_{ij}\cdot Pd_{ij}\cdot Pl_{ij}
\end{equation}

\noindent In the sensitive ant model the ants with high pheromone level are choosing the next node based on (\ref{eq:3}) from the neighborhood $J$ of node $j$.
\begin{equation}\label{eq:3}
j=argmax_{u\in J_{ik}}\{(\varphi_{iu}\cdot\eta_{iu})^\alpha\cdot(\frac{1}{D_{iu}})^\beta\}.
\end{equation}

\noindent $\varphi_{ij}$is the pheromone intensity between the source node $i$ and destination node $j$; The normalized value is  the difference between total and actual value of the performance parameters.  The performance value is used to compute the transition probabilities in a route. The link being active or dead in a tour taken by an ant is incorporated in the pheromone. The pheromone is globally updated~\cite{dorigo} following each complete tour by ant system with the update rule (\ref{eq:4}) is following~\cite{Muraleedharan}.
\begin{equation}\label{eq:4} 
\varphi_{ij}(t)=\rho(\varphi_{ij}(t-1))+\frac{Q}{D_t\cdot \eta_t}
\end{equation}
\noindent where $D_t$ is the total distance traveled by ants during the current tour. The trails formed by the ant is dependent on the link factor.  The tabu list includes now updated values of the energy available in the nodes for a particular sub-optimal route with high reach-ability. A run of the algorithm returns the valid path of the wireless network. That is how the information in {\it WSN} is re-routed. Termination criteria is given by a given number of iterations.

\noindent {\bf Discussions.} Sensitive ants with lower pheromone level are able to explore the wireless network and the ants with high pheromone level intensively exploit the promising search regions already identified in the network. The ant’s behavior emphasizes search intensification. The ants ``learn'' during their lifetime and are capable to improve their performances. That is how they modifies their level of sensitivity: the PSL value increase or decrease based on the search space topology encoded in the ant’s experience.

\noindent The introduced model {\em  Sensitive Ant Algorithm for Denial Jamming Attack on WSN} seems to improve the jamming attack detection and re-routing in wireless sensor network.
\section{Conclusions} The paper shows the main jamming attacks and several countermeasure. It is introduced a new {\it Sensitive Ant Denial Jamming} on {\it Wireless Sensor Network} based on ant system as mobile-agents class. In general mobile agent techniques including ant models proved to have a medium defense effectiveness, a medium cost but a good compatibility with existing software. The introduced sensitive model brings a new feature that improves the reactions of agents in the network in case of jamming attacks and redirect the information also to the processing element in order to re-routing information.
\section*{Acknowledgement.}\noindent This research is supported by Grant PN II TE 113/2011, New hybrid metaheuristics for solving complex network design problems, funded by CNCS Romania.

\end{document}